\documentclass[prl]{revtex4}
\usepackage{graphicx,epstopdf,amsmath}
\usepackage{setspace}

\begin{document}
\newcommand{\mpl}{M_{Pl}^2}
\newcommand{\rt}{R_{T}}
\newcommand{\rf}{R_F}
\newcommand{\sint}{\sin(\Theta)}
\newcommand{\sit}{\sin(\Theta)}
\newcommand{\sints}{\sin^2(\Theta)}
\newcommand{\sintc}{\sin^3(\Theta)}
\newcommand{\ct}{\cos(\Theta)}
\newcommand{\cts}{\cos^2(\Theta)}
\newcommand{\ctc}{\cos^3(\Theta)}
\newcommand{\be}{\begin{equation}}
\newcommand{\ee}{\end{equation}}
\newcommand{\tm}{\widetilde{\mathcal{M}}}
\newcommand{\pit}{\frac{\pi}{2}}
\newcommand{\gpi}{\Theta > \pit}
\newcommand{\lpi}{\Theta < \pit}
\newcommand{\tr}{\widetilde{R}}
\newcommand{\tl}{\widetilde{\Lambda}}
\newcommand{\ds}{ \delta(\sigma-\sigma_{wall})}
\newcommand{\psq}{\phi_{ab}\phi^{ab}}
\title{Horizons and Tunneling in the Euclidean False Vacuum}
\author{Kate Marvel and Neil Turok }
\address{Department of Applied Mathematics and Theoretical Physics, Cambridge University, Wilberforce Road, Cambridge CB3 0WA  }
\date{\today}
\begin{abstract}
In the thin-wall approximation, the decay of a gravitating false vacuum to a lower-energy state is affected by the cosmological horizon structure in both spaces.  The nucleation radius of a bubble of true vacuum depends on the surface tension of its boundary and equals the false vacuum cosmological horizon at a critical tension.  We argue that there is no tunneling instanton solution beyond the critical tension and  argue that there is therefore a bound on allowed membrane tension in theories which rely on semiclassical tunneling to relax the cosmological constant.
\end{abstract}
\maketitle

\section{Introduction}
The cosmological constant problem \cite{Weinberg:1988cp} is one of the most intractable in modern physics.  There is still no fully convincing explanation for why the observed vacuum density is over 120 orders of magnitude less than the predicted zero-point energy of the standard model, nor for why it has come to dominate the Universe so recently.   

A tantalizing possibility is that the cosmological vacuum density is not constant, but changes in accordance with some more fundamental theory.  One of the first attempts to make the ``constant" dynamical was proposed by Abbot \cite{Abbott:1984qf}, who proposed that the source of dark energy could be a scalar field in a potential with an infinite set of equally spaced metastable local minima.  If the field starts high on the potential, it will roll down until it reaches a point where the barrier between minima becomes significant.  After this point the field may proceed toward the potential absolute minimum by tunnelling between metastable states.  The cosmological constant is therefore relaxed as the field jumps slowly down the potential.  

Another dynamical scenario, first proposed by Brown and Teitelboim \cite{Brown:1988kg}, \cite{Brown:1987dd}, involves the spontaneous creation of membranes within a de Sitter background.  Each membrane, described by an appropriate instanton solution, raises or lowers the observed value of the cosmological constant by an integer multiple of some fundamental energy density \cite{Hawking:1984hk}.   This is because the membranes are two-dimensional charged objects, so they source a three-form potential $A_3$ and an antisymmetric four-form flux $F_4=d A_3$.  The action for gravity and this field strength tensor is 
\be
S=\int d^nx \sqrt{-g} \left(\frac{\mpl}{2} R - \Lambda_{bare} - \frac{1}{2 \cdot 4!} F_{4}^2\right)
\ee
where the bare cosmological constant $\Lambda_{bare}$ is assumed to be negative and large.  The four-form equation of motion is then $\partial_{a} (\sqrt{-g} F^{a b c d })=0$, so the incorporation of the $F_4^2$ term in the action mimics the effect of a postive cosmological constant.  The effective observable is therefore $\Lambda_{bare} + C$, where $C \propto F_4^2 = F_{abcd}F^{abcd}$ is a constant.  The spontaneous nucleation of membranes is a fairly well-understood non-perturbative effect analogous to the tunneling used in the Abbott model \cite{Giddings:1987cg}.  Each nucleated membrane gives rise to a four-form flux, and so acts as a boundary between regions of different flux and different observed $\Lambda$.

These proposals, however, both suffer from serious problems that must be dealt with in any dynamical dark energy model.  First, with a changing cosmological constant, it is extremely difficult to populate the Universe with some form of matter or radiation.  Every time the Universe settles into some new potential minimum, the resulting de Sitter phase will inflate away any matter and radiation.  This ``empty Universe" problem means that by the time the cosmological constant reaches the observed value, almost all matter and radiation will have been inflated away.  Second, these models require an energy spacing $\delta \Lambda$ infinitesimal compared with all known scales of physics; this is the so-called ``gap problem".
 
Several authors have attempted to resolve the gap problem by modifying the Brown-Teitelboim proposal  \cite{Bousso:2000xa}, \cite{Feng:2000if}.  In these scenarios, the cosmological constant is dynamically neutralized by different species of four-form field strength produced by wrapping compact dimensions with flux, a process that can be executed in a number of different ways.  These field strengths are quantized in integer multiples of some basic unit, giving rise to a nonzero energy density.  The measured cosmological constant is then
$$
\Lambda= \Lambda_{bare} +\sum_{i=1}^{N} c_i q_i^2
$$
where the first term is the bare cosmological constant, assumed to be negative.  In the second term, the integer charges $q_i$ arise from the geometry of the compactification manifold, and $N$ is the number of wrapped cycles present in the compactified space.  Bousso and Polchinski found that the ``gap problem" could be evaded in an M-theory context by incorporating multiple four-forms, perhaps resulting from wrapping seven-forms about three-cycles of the compact space.  If there is more than one non-trivial three-cycle, then naturally more than one type of four-form can arise from the wrapping.  It may then be possible to find some combination of different fluxes that can neutralize the cosmological constant in a natural way.  As with the Brown-Teitelboim model, the relaxation of the cosmological constant is achieved by the spontaneous creation of membranes that separate regions of differing flux.   However, these proposals do not directly address the empty Universe problem of the dynamical relaxation scenario.

If a dynamical model is to solve \emph{both} problems, however, it must propose a radical new view of cosmic history.  Steinhardt and Turok propose tosolve the empty Universe and gap problems by embedding Abbott's dynamical vacuum energy model in a cyclic model \cite{Steinhardt:2006bf}.  Because matter and radiation in this scenario are created by a collision of two orbifold planes \cite{Khoury:2001zk}, the empty Universe problem is solved if the collision timescale is much shorter than the relaxation time.  Additionally, since the cyclic model does not require an inflationary period to generate a scale-invariant spectrum of density perturbations \cite{Steinhardt:2002ih}, there is no constraint placed on the relaxation by an inflationary epoch.  

Finally, the cosmological constant could be non-dynamical, and its explanation anthropic ~\cite{Weinberg:1987dv}.  Such arguments depend on the availability of a large solution space with a well-defined measure.  Recent progress in string theory suggests that the space of solutions can be parametrized by moduli which act, at low energies, like scalar fields ~\cite{Susskind:2003kw}.  The ``landscape" potential that results from changing the values of these fields appears to offer a wide range of discrete vacua, each with a different cosmological constant.  The goal of the anthropic approach is to use this huge space of solutions to explain why the laws of physics are as observed.  There does not, as yet, appear to be a universally agreed-upon measure with which to assign probabilities to these scenarios, although several candidates have been proposed \cite{Gibbons:1986xk}, \cite{Gibbons:2006pa}.  Even within this framework, many anthropic arguments rely on the relative stability of the vacua, and it is important to understand the mechanisms for changing vacuum state \cite{HenryTye:2006tg}.

\subsection{False Vacuum Decay}
In the vacuum, the classical solution to Einstein's equations with a positive cosmological constant is de Sitter space. A de Sitter vacuum, however, is a metastable one; there are  at least two well-known processes by which it may decay to a state of lower energy.  First, Ginsparg and Perry \cite{Ginsparg:1982rs} have shown that de Sitter space is unstable to the nucleation of black holes.  This effect, similar in nature to the decay of hot flat space, is suppressed at small $\Lambda$ and we shall not consider it further.  The second decay process involves the evolution of a de Sitter universe into another spacetime of maximal symmetry, possibly a de Sitter space of lower cosmological constant or even flat space.  If an underlying theory allows multiple de Sitter-like vacua, it is possible for a bubble of lower vacuum energy to form inside the outside, ``false" vacuum.  This ''true" vacuum bubble then expands at constant acceleration.  It is this process that we will study in detail in the following sections.

%The lifetime of any given de Sitter vacuum is of paramount importance in dynamical model-building.  In the subsequent sections, we argue that the presence of this horizon, apparent as the equator of the Euclidean false vacuum, affects the nature of the solutions, and therefore the semiclassical lifetime of any given vacuum.
\subsection{Coleman-DeLuccia instantons}
\begin{figure}[htbp]
\begin{center}
\includegraphics[scale=.3]{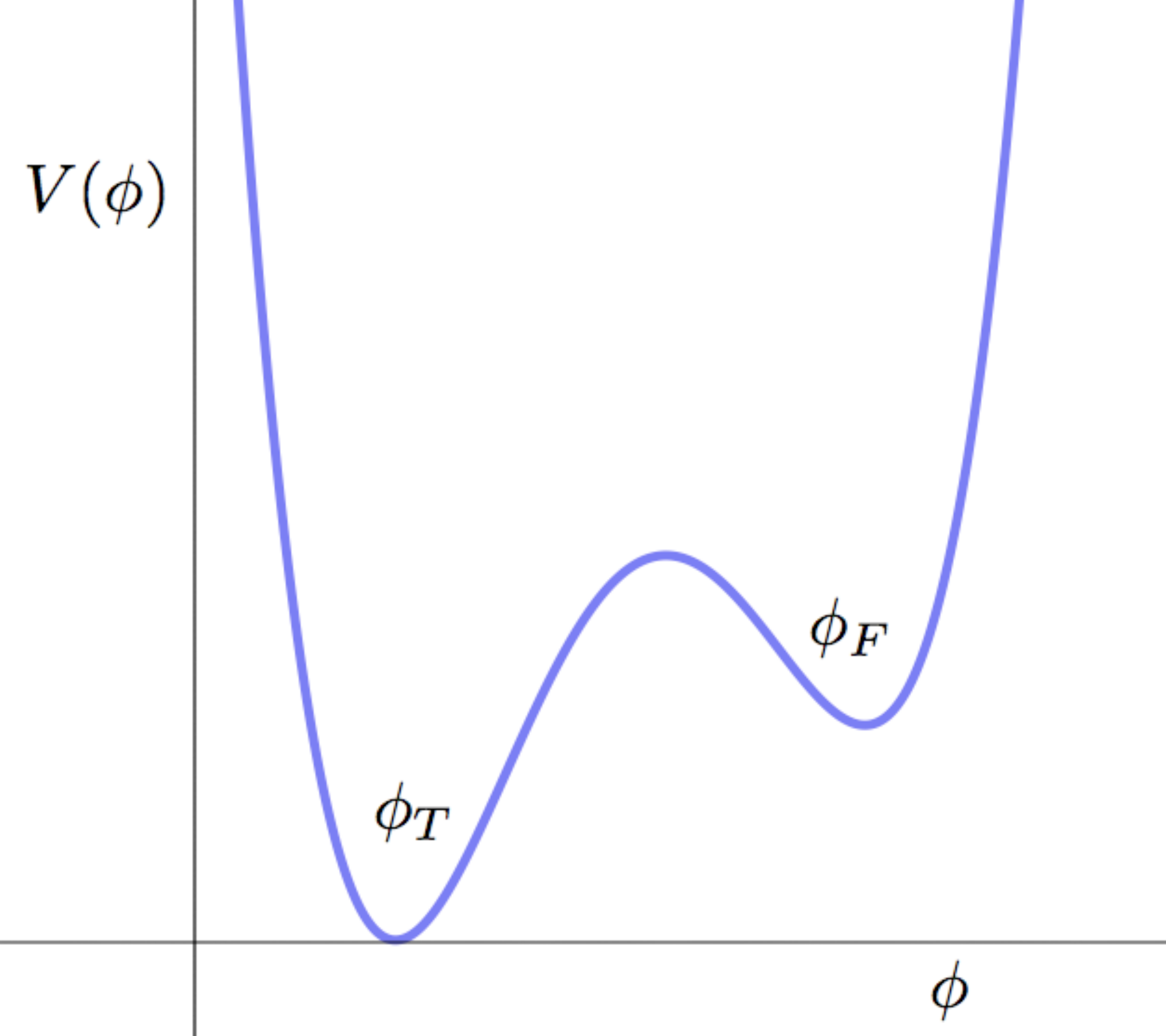}
\caption{The potential for tunneling.}
\label{cdpotential}
\end{center}
\end{figure}

An elegant description of vacuum decay was developed by Sidney Coleman and collaborators \cite{Coleman:1977py}, \cite{Callan:1977pt}, \cite{Coleman:1980aw}, who consider a field in a potential with both metastable and true minima, as shown in Figure \eqref{cdpotential}.

By analogy with the WKB method in quantum mechanics, a tunneling solution is a classical stationary point of the Euclidean action, i.e. an instanton.  The semiclassical process is then interpreted as the materialization of a bubble of true vacuum of finite radius inside the outer false vacuum space.  This  bubble nucleation radius $r_0$ is determined by the spacing  $\epsilon$ between true and false vacua and by the thickness $S_1 = \int _{\phi_{f}}^{\phi_{T}} d\phi \; \left(2 \left [ V(\phi)-V(\phi_F)\right] \right)^{-1/2}$ of the transition region. Its subsequent evolution is then determined by analytic continuation to Lorentzian signature.   If the spacing $\epsilon$ is small,  then the thickness of the transition zone between false and true vacua can be made much smaller than the bubble radius; this is the so-called ``thin-wall" approximation.  

The inclusion of gravity modifies the bubble radius, as gravitational effects must be balanced against volume and surface energy.  It is convenient to parameterize the vacuum by the de Sitter radius, related to the cosmological constant by $R_{dS}=\sqrt{3 \mpl /\Lambda}$. For the decay of de Sitter space with radius $R_{dS}$ into flat space, Coleman and DeLuccia find the nucleation radius to be
\be
\frac{r_{nuc}}{R_{dS}}=\frac{r_0/R_{dS}}{1+(r_0/2 R_{dS})^2}
\ee
where $r_0$ is the bubble radius in the absence of gravity.   The gravitating bubble radius $r_{nuc}$ therefore appears to increase with $r_0$ up to the horizon radius $R_{dS}$, after which it appears to decrease.  The behavior of the solution at this point has been the subject of some controversy: some authors \cite{Steinhardtunpub}, \cite{Maloney:2002rr} advocate discarding all solutions beyond a certain $r_0= R_{dS}$ .  

\begin{figure}[htbp]
\begin{center}
\includegraphics[scale=.4]{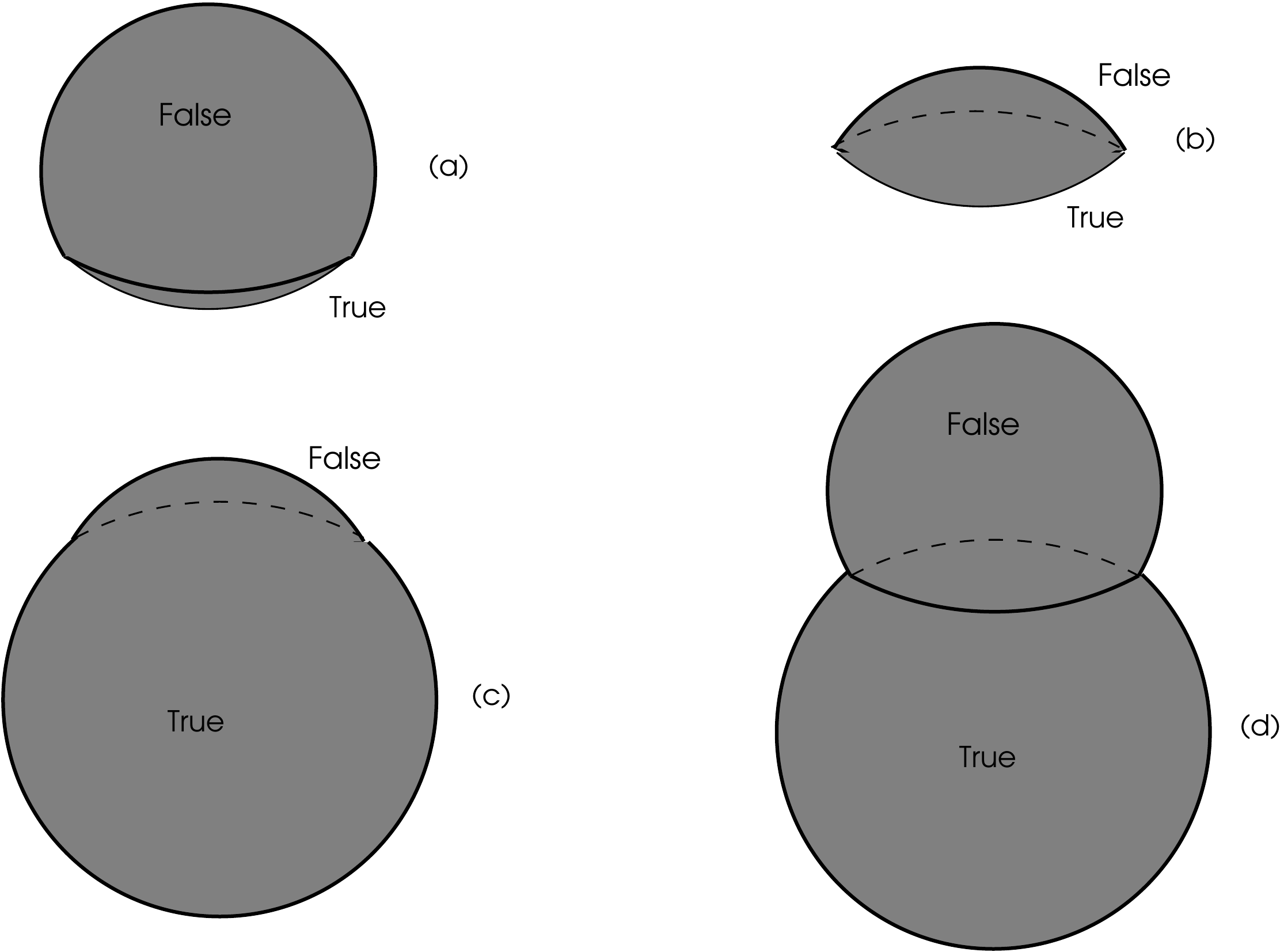}
\caption{Four different instantons corresponding to a single nucleation radius $r_{nuc}$.}
\label{fourclasses}
\end{center}
\end{figure}

The difficulty stems from the fact that Euclidean de Sitter space has the topology of a four-sphere, and a given value of $r_{nuc} <  R_{dS}$ is achieved \emph{twice} on each $S^4$.  For any given nucleation radius, we shall see that there are four possible instanton configurations \cite{Steinhardtunpub}, as shown in Figure \eqref{fourclasses}.   Each has a different physical interpretation.  In configurations (a) and (d), the true vacuum bubble nucleates well below the  equatorial radius of the false vacuum, whereas  in the configurations (b) and (c)  the true vacuum swallows a much larger proportion of the false vacuum.   Likewise, (a) and (d) can be differentiated by the amount of true vacuum nucleated in the false -- in (a) the true vacuum region fits comfortably in the false, whereas in (d) it does not.

In the subsequent sections we will break this degeneracy by showing that only the configuration labeled (a) possesses a negative mode and obeys the relevant stress-energy conservation conditions.  It is therefore the only solution that can sensibly be interpreted as a tunneling one.

\section{Action}
To prove a stationary point of the Euclidean action corresponds to a tunneling solution,  we consider the spectrum of the second variational derivative of $S_{E}$ at the point.  Negative eigenvalues give an imaginary contribution to the energy \cite{Coleman:1987rm}, rendering the state unstable, and the existence of negative modes is crucial to the interpretation of an instanton as a tunneling solution \cite{Gratton:1999ya}, \cite{Gratton:2000fj}.  We argue that for a stationary point of the Euclidean action to correspond to a semiclassical, $O(4)$ symmetric tunneling process, it must have instabilities, manifest as negative modes of the second variational derivative.   The existence of these negative modes has sparked considerable interest; see \cite{Tanaka:1992zw}, \cite{Tanaka:1999pj}, \cite{Lavrelashvili:1999sr}. Our analysis rests on the symmetry of the problem; we assume both false and true Euclidean spaces are pure de Sitter, and that the wall separating them is a perfect three-sphere.   This assumption, used frequently in the literature (see, e.g. \cite{Rubakov:1999ir} and \cite{Brown:2007sd}), allows us to reduce the space of solutions to a one-parameter family characterized by the angular coordinate of the wall.   
To begin, we dispense with the scalar field and the potential altogether.  The Euclidean action for a tunneling solution such as that described in Fig. \eqref{fourclasses} is then described purely by the Einstein-Hilbert action for gravity 
\be
\frac{\mpl}{2}\int _{\widetilde{\mathcal{M}}}\sqrt{-g} (R-2 \Lambda)d^4 x .\label{EinsteinHilbert}
\ee
Here, the instanton manifold $\tm$ is the connected sum of two submanifolds $\mathcal{M}_T$ and $\mathcal{M}_F$ representing true and false vacua respectively.  We now show that, for given vacuum manifolds, this can be written in terms of a single parameter, namely the tension on the bubble wall itself.

%$$
%\tm =(\mathcal{M}_T \backslash \Sigma) \# (\mathcal{M}_F \backslash \Sigma).
%$$
In the thin-wall approximation, the boundary $\Sigma$ between the true and false vacua is a round three-sphere.  By defining a vector $n_a$ everywhere normal to this surface, we can relate the induced metric on $\Sigma$, denoted by $h_{ab}$, to the metrics $g_{ab}$ defined on $\mathcal{M}_T$ and $\mathcal{M}_F$:
\be
g_{ab} = h_{ab} + n_a n_b.\label{induced}
\ee
Because both true and false spaces are in vacuo,  only this boundary will contribute to the energy of the configuration.  In the thin-wall approximation, the energy-momentum tensor on $\tm$ receives a contribution from $\Sigma$ and can be written as
 \be
T_{ab} = \frac{T \ds}{3} (g_{ab} - n_a n_b) 
\ee  
where the delta function reflects the jump across the boundary between true and false vacua \cite{Blau:1986cw}.
Taking the trace of the Einstein equations allows us to write the three-dimensional Ricci scalar on $\Sigma$ as 
$$^3 R = -\mpl T,$$
where $T$ can be regarded as the tension of the bubble wall.   
The second fundamental form of this boundary surface, denoted $K_{ab}$, is calculated by taking the Lie derivative of the induced metric on $\Sigma$ along the normal vector:
$$K_{ab} = \frac{1}{2} \mathcal{L}_{n} h_{ab}.$$
The Einstein-Hilbert action \eqref{EinsteinHilbert} contains terms linear in second derivatives of the metric, which can be integrated by parts, allowing us to rewrite the Einstein-Hilbert action on $\tm$ in terms of first derivatives \cite{Gibbons:1976ue} as 
\begin{eqnarray}
S_{grav}=\frac{-\mpl}{2} \int_{M} d^4 x \; \sqrt{-g_f} (R_{f}-6 \rf^{-2}) -
\frac{\mpl}{2} \int_{M} d^4 x \; \sqrt{-g_t} (R_{t}-6 \rt^{-2}) - \nonumber \\
\mpl  \int_{\partial M} d^3 x \; \sqrt{-h} K_{f} -
\mpl  \int_{\partial M} d^3 x \; \sqrt{-h} K_{t}+
T  \int_{wall} d^3x \; \sqrt{-h}
\label{gravac}
\end{eqnarray}
Here, $g_t$ and $g_f$ are the metric determinants in the true and false vacua, respectively, while $h$ is the induced metric determinant on the boundary separating false vacuum from true.  $K_t$ and $K_f$ are the  traces of $K_{ab}$ calculated on either side.  We are interested in evaluating the Euclidean action, defined as the imaginary portion of the gravitational action evaluated along a contour on which the true vacuum bubble evolves from zero size.  When exponentiated, this quantity can be interpreted as a tunneling rate per unit volume if negative modes are present, as explained in \cite{Coleman:1980aw}.

\begin{figure}[htbp]
\begin{center}
\includegraphics[scale=.4]{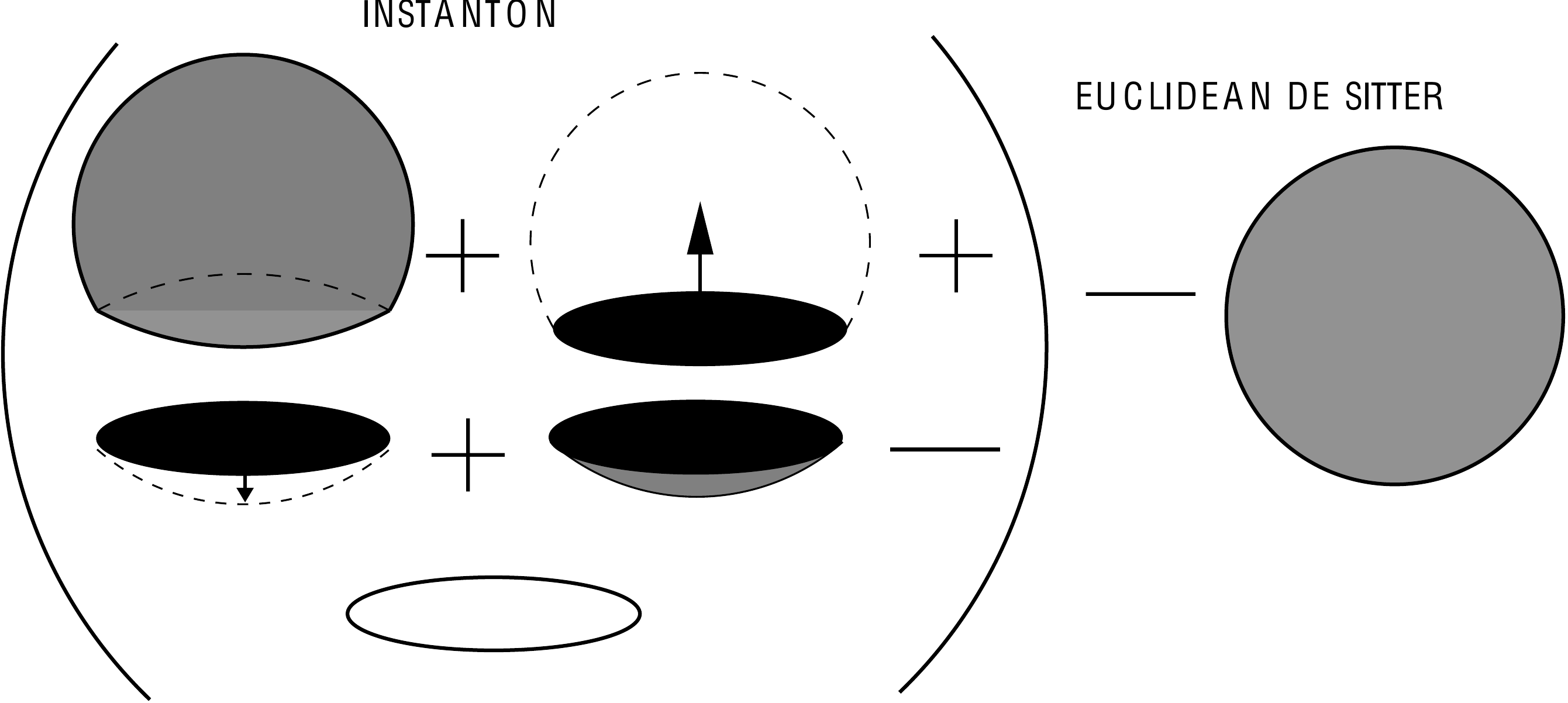}
\caption{{\bf Instanton solution for false vacuum decay.}}
\label{goodinstanton}
\end{center}
\end{figure}

For simplicity, we will work in Gaussian normal coordinates in which the boundary wall is located at $\sigma=\sigma_{wall}$. Each Euclidean de Sitter space has the topology of a four sphere with radius $R_{dS}$, so the metric in both true and false spacetimes is 
\be 
ds^2 = d\sigma^2 +b^2(\sigma) d\Omega_{(3)}^2\label{EucMetric}
\ee
with $b=R_{dS} \sin(\sigma/R_{dS})$.  The instanton described by Fig. \eqref{fourclasses} is therefore constructed by cutting the four-sphere representing the false vacuum along a line $\sigma=\sigma_{wall}$ and ``gluing" it to the true vacuum, a four-sphere cut along the same line.  The contributions to the action come from the volumes of the spherical caps, the energy of the three-dimensional boundary itself, and the gravitational effects induced on the boundary by both sides. 
This is depicted schematically in Figure \eqref{goodinstanton}.  

The metric $\widetilde{g_{ab}}$ on $\tm$ is therefore defined as
\be
b(\sigma) = \left\{ \begin{array}{l}
\rf \sin(\sigma/\rf) \; \; \mathrm{if}\; \; 0 \leq \sigma \leq \sigma_{wall} \\
\rt \sin(\sigma/\rt)\; \; \mathrm{if}\; \; \rt \sin^{-1}(\rf/\rt \sin(\sigma_{wall}/\rf))\leq \sigma \leq \pi
\end{array}\right. \label{metric}
\ee
so that the metric is continuous at the wall.   In these coordinates  the extrinsic curvature in both true and false vacua can be written in terms of the induced metric $h_{ab} $:
\be
K_{ab} = -\Gamma_{ab}^{0}= \frac{1}{2} \frac{\partial g_{ab}}{\partial \sigma} = b b' h_{ab}\label{Kh}
\ee
In these coordinates, the Euclidean action becomes
\begin{center}
\begin{eqnarray}
S_E=T\int \delta(\sigma-\sigma_{boundary})d\sigma \; d\Omega_{(3)}^2-\frac{\mpl}{2} \int_{false} 6(b+b b'^2 - b^3 R^{-2}) d\sigma \; d\Omega_{(3)}^2 \nonumber \\ -\frac{\mpl}{2}
 \int_{true}  6(b+b b'^2 - b^3 R^{-2}) d\sigma \; d\Omega_{(3)}^2. \label{prettyaction}
\end{eqnarray}
 \end{center}
 
On shell, the function $b(\sigma)$ satisfies the constraint 
 \be
 1-\frac{b^2}{R^2} = b'^2\label{Hamiltonian},
 \ee
which can be used to eliminate the $b'^2$ terms.  We can then impose 
$\frac{\delta S}{\delta b} =0$ to derive the Israel matching condition \cite{Israel:1966rt}
\be
\left [ \frac{b'}{b} \right ]^F_T=-\frac{T}{2 \mpl} \label{Israel}
\ee
relating the surface energy density of the bubble wall to the jump in the extrinsic curvature between false vacuum and true.   We now proceed to use this condition to investigate the stationary points of the action.
\subsection{Stationary Points}
\begin{figure}[htbp]
\begin{center}
\includegraphics{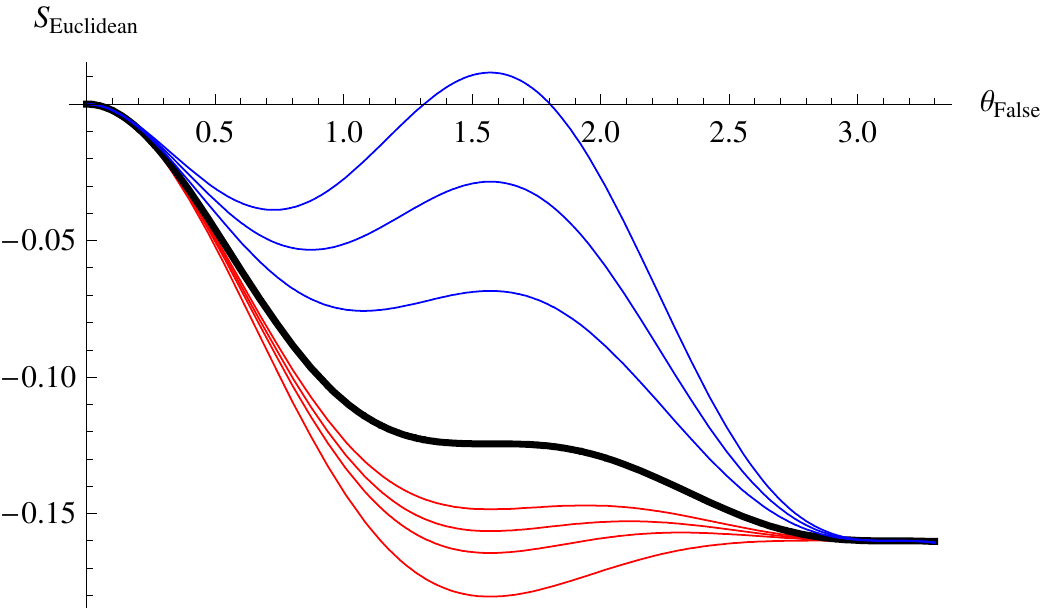}
\caption{Euclidean action as a function of bubble nucleation point.  The actions for tensions less than the critical tension are shown in red, those for supercritical tensions are shown in blue.  The action corresponding to the critical tension is represented by the thick black line.}
\label{actionplot}
\end{center}
\end{figure}

 Defining $\Theta =\sigma \rf$, $\alpha=\frac{\rf T }{2 \mpl}$, and $\beta = \frac{\rt}{\rf}$,  and integrating over all coordinates, the Euclidean action \eqref{prettyaction} becomes

\begin{eqnarray}
\frac{S_E}{2 \pi^2}=2 \mpl \rf^2\left(-1 - \beta^2 + \cos^3(\Theta) + \alpha \sin^3 (\Theta) + \beta^2 \left(1- \frac{\sin(\Theta)^2}{\beta^2} \right)^{3/2} \right).\label{theaction}
\end{eqnarray}
and imposing $\delta S=0$, we see that for any tension $T$, there is a Euclidean solution corresponding to the bubble wall location
\be
\Theta=\sin^{-1}\left( \frac{ 2 \alpha \beta^2}{\sqrt{1+2 (\alpha^2 -1) \beta^2+(\alpha^2 +1) \beta^4}}\right)\label{stationary}
\ee
This allows for a more complete interpretation of the four instanton classes in Fig \eqref{fourclasses}.    In classes (a) and (d), the coordinate $\Theta > \pit$,  and in (b) and (c) $\lpi$.  If the de Sitter radii of the true and false vacua are fixed, the parameter $\alpha$ is the sole determinant of the coordinate $\Theta$.  
 For small $\alpha$,then, there is a solution corresponding to a wall located  at $\Theta > \frac{\pi}{2}$.  For larger tensions the solution is pushed closer and closer to the equator of the false vacuum four sphere, with equality when $1+\alpha^2\beta^2 = \beta^2$.  In dimensional units, this corresponds to a critical tension on the bubble wall
\be 
T_{crit}= \frac{2 \mpl \sqrt{\rt^2-\rf^2}}{\rt \rt}.\label{Tcrit}.
\ee
  For tensions exceeding this critical value, the only possible nontrivial configuration involves a bubble wall at $\Theta < \frac{\pi}{2}$.    The effect of increasing tension on the critical points of the action is shown in Figure \eqref{actionplot}.

It appears that for any tension, the Euclidean action has three other stationary points.  However, the solution $\Theta =0$ satisfies the Israel matching condition \eqref{Israel} only if $T=0$, while the solution $\Theta =\pi$ is a solution only if $T=\infty$.  The horizon-size bubble $\Theta = \pi/2$ satisfies Israel matching only for the critical tension $T_{crit}$.  Israel matching also allows us to rule out two instanton classes in Figure \ref{fourclasses}.  If we assume that the wall tension $T$ is positive, than only configurations $(a)$ and $(b)$ in Figure \eqref{fourclasses} satisfy this constraint.

We shall now look for negative modes, the existence of which will depend wholly on this bubble wall coordinate $\Theta$.

\subsection{Second Derivative}
\begin{figure}[htbp]
\begin{center}
\includegraphics{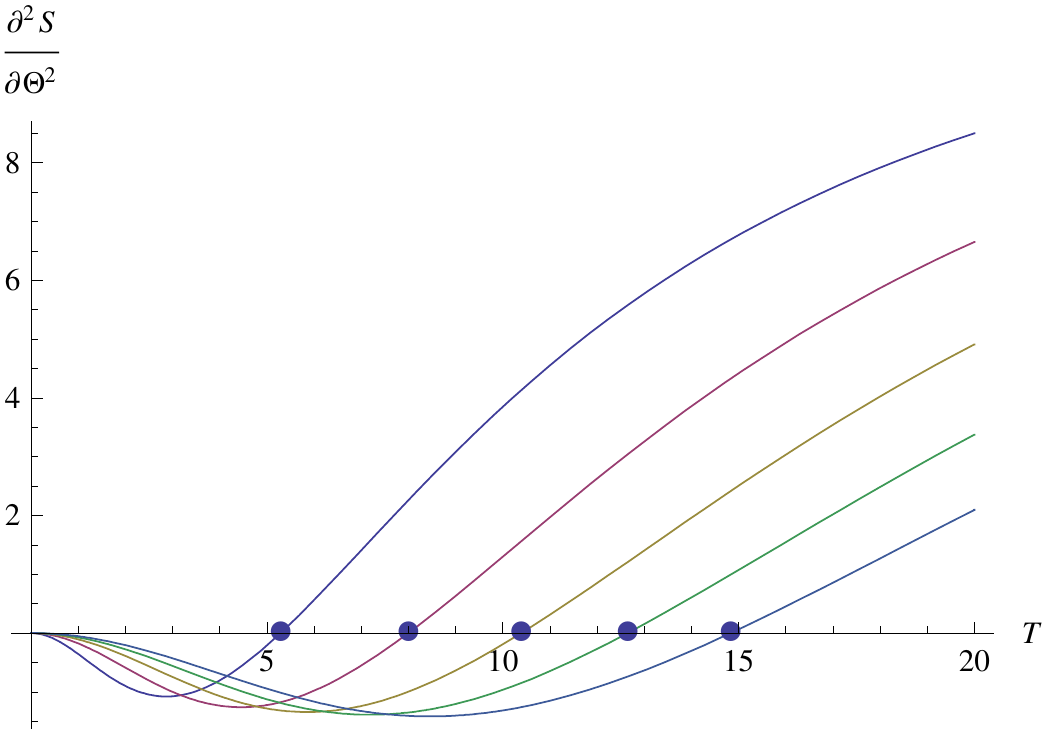}
\caption{The second derivative evaluated at the critical point for multiple choices of the parameters $(\rf,\rt)$.  The critical tension for each choice is plotted as a dot.} \label{ddplot}
\end{center}
\end{figure}

If we define $\mathcal{S}=\frac{S_E}{4 \pi^2 \rf \mpl}$, the second variational derivative reduces to the second derivative with respect to $\Theta$,
\be 
\frac{\delta \mathcal{S}^2}{\delta \Theta^2} = -\frac{1}{8 \beta \sqrt{\beta^2 - \sin^2(\Theta)}} 3 (1+2 (2 \beta^2 - 1) \cos(2 \Theta) + 3 \cos(4\Theta)+2 \beta \sqrt{\beta^2-\sin^2(\Theta)} (\ct + 3 \cos (3 \Theta) + \alpha(\sint-3\sin(3 \Theta)))
\ee
and is zero when $1+\alpha^2\beta^2 = \beta^2$.  The second variational derivative is  therefore negative for $T<T_{crit}$ and positive for $T>T_{crit}$.  Figure \eqref{ddplot} shows the second derivative evaluated at the critical point for various choices of parameters.

%$$\frac{S_{E}}{2 \pi} =2 \mpl \left(  \rf^2 -\rf^2 \cos^3(\psi) + \frac{\rf^2 T \sin^3(\psi)}{2 \mpl} +\rt^2 \left (1-\left(\frac{\rf}{\rt} \right)^2 \sin^2(\psi) \right )^{3/2} \right)$$

But because tensions greater than $T_{crit}$ force the bubble outside the false vacuum horizon, we conclude that negative modes are present only for configurations in which $\Theta > \pit$; that is, there is a maximum tension allowable for membranes which serve as boundary manifolds in semiclassical tunneling.
\section{Summary}
The four instanton classes in Fig \eqref{fourclasses} are summarized in Table \eqref{sumtable}.
\begin{table}[h]
\begin{center}
\begin{tabular}{|c| c| c |  c| c|}
\hline
Class & $\Theta_{false}$ & $\Theta_{true}$ & Israel Matching Satisfied & Negative Mode \\ \hline
a& $\gpi $&$\lpi$ &Yes & Yes\\ \hline
b& $\lpi i $&$\lpi$ &Yes & No\\ \hline
c& $\lpi $&$\lpi$ &No & No\\ \hline
d& $\gpi $&$\gpi$ &No &Yes\\ \hline
\end{tabular}
\caption{Summary of instanton results.}\label{sumtable}
\end{center}
\end{table}
If we assume only positive tension branes are allowable, then the only solution that both satisfies Israel matching and has a negative mode is class (a).  
We believe this resolves a long-standing issue in vacuum decay and illuminates the precise influence of cosmological horizons in both vacua on the tunneling solution. 

\section{Acknowledgements}
We thank Paul Steinhardt for useful discussions which stimulated this work, and for sharing \cite{Steinhardtunpub} with us.  KM thanks Malcolm Perry and Daniel Wesley for helpful conversations and Trinity College Cambridge for support.

\bibliographystyle{h-physrev3.bst}
\bibliography{Euclidean}

\end{document}